\documentclass[twocolumn]{aastex63}

\usepackage{graphicx}
\usepackage{amsmath}
\usepackage{amssymb}
\usepackage{xspace}

\newcommand{\rr}[1]{#1}

\newcommand{\ha}{\texorpdfstring{\ensuremath{\mathrm{H}\alpha}}{Halpha}\xspace}
\newcommand{\hb}{\texorpdfstring{\ensuremath{\mathrm{H}\beta}}{Hbeta}\xspace}

\newcommand{\ergs}{erg\,s$^{-1}$\xspace}
\newcommand{\ergsa}{erg\,s$^{-1}$\,\AA$^{-1}$\xspace}
\newcommand{\ergscm}{erg\,s$^{-1}$\,cm$^{-2}$\xspace}
\newcommand{\ergscma}{erg\,s$^{-1}$\,cm$^{-2}$\,\AA$^{-1}$\xspace}

\received{---}
\accepted{---}

\submitjournal{ApJ}

\shorttitle{HST observations of 6 GASP RPS galaxies}
\shortauthors{Gullieuszik et al.}

\begin{document}

\title{UV and \ha\ HST observations of 6 GASP jellyfish galaxies}

\author[0000-0002-7296-9780]{Marco Gullieuszik}
\affiliation{INAF-Osservatorio astronomico di Padova, Vicolo Osservatorio 5, 35122 Padova, Italy}

\author[0000-0002-3818-1746]{Eric Giunchi}
\affiliation{INAF-Osservatorio astronomico di Padova, Vicolo Osservatorio 5, 35122 Padova, Italy}
\affiliation{Dipartimento di Fisica e Astronomia, Universit\`a di Padova, Vicolo Osservatorio 3, 35122 Padova, Italy}

\author[0000-0001-8751-8360]{Bianca M. Poggianti}
\affiliation{INAF-Osservatorio astronomico di Padova, Vicolo Osservatorio 5, 35122 Padova, Italy}

\author[0000-0002-1688-482X]{Alessia Moretti}
\affiliation{INAF-Osservatorio astronomico di Padova, Vicolo Osservatorio 5, 35122 Padova, Italy}

\author[0000-0002-9136-8876]{Claudia Scarlata}
\affiliation{Minnesota Institute for Astrophysics, School of Physics and Astronomy, University of Minnesota, 316 Church Street SE, Minneapolis, MN 55455, USA}

\author[0000-0002-5189-8004]{Daniela Calzetti}
\affiliation{Department of Astronomy, University of Massachusetts, 710 N. Pleasant Street, LGRT 619J, Amherst, MA 01002, USA}

\author[0000-0002-4382-8081]{Ariel Werle}
\affiliation{INAF-Osservatorio astronomico di Padova, Vicolo Osservatorio 5, 35122 Padova, Italy}

\author[0000-0001-8600-7008]{Anita Zanella}
\affiliation{INAF-Osservatorio astronomico di Padova, Vicolo Osservatorio 5, 35122 Padova, Italy}

\author[0000-0002-3585-866X]{Mario Radovich}
\affiliation{INAF-Osservatorio astronomico di Padova, Vicolo Osservatorio 5, 35122 Padova, Italy}

\author[0000-0002-6179-8007]{Callum Bellhouse}
\affiliation{INAF-Osservatorio astronomico di Padova, Vicolo Osservatorio 5, 35122 Padova, Italy}

\author[0000-0002-4158-6496]{Daniela Bettoni}
\affiliation{INAF-Osservatorio astronomico di Padova, Vicolo Osservatorio 5, 35122 Padova, Italy}

\author[0000-0001-9575-331X]{Andrea Franchetto}
\affiliation{INAF-Osservatorio astronomico di Padova, Vicolo Osservatorio 5, 35122 Padova, Italy}

\author[0000-0002-7042-1965]{Jacopo Fritz}
\affiliation{Instituto de Radioastronomia y Astrofisica, UNAM, Campus Morelia, AP 3-72, CP 58089, Mexico}

\author[0000-0003-2150-1130]{Yara L. Jaff\'e}
\affiliation{Instituto de Física y Astronomía, Universidad de Valparaíso, Avda. Gran Bretaña 1111 Valparaíso, Chile}

\author[0000-0003-3255-3139]{Sean L. McGee}
\affiliation{School of Physics and Astronomy, University of Birmingham, Birmingham B15 2TT, UK}

\author[0000-0003-2589-762X]{Matilde Mingozzi}
\affiliation{Space Telescope Science Institute, 3700 San Martin Drive, Baltimore, MD 21218, USA}

\author[0000-0002-0838-6580]{Alessandro Omizzolo}
\affiliation{Specola Vaticana, 00120, Vatican City State}
\affiliation{INAF-Osservatorio astronomico di Padova, Vicolo Osservatorio 5, 35122 Padova, Italy}

\author[0000-0002-8710-9206]{Stephanie Tonnesen}
\affiliation{Center for Computational Astrophysics, Flatiron Institute, 162 5th Avenue, New York, NY 10010, USA}

\author[0000-0001-9022-8081]{Marc Verheijen}
\affiliation{Kapteyn Astronomical Institute, University of Groningen, Landleven 12, NL-9747 AV Groningen, the Netherlands}

\author[0000-0003-0980-1499]{Benedetta Vulcani}
\affiliation{INAF-Osservatorio astronomico di Padova, Vicolo Osservatorio 5, 35122 Padova, Italy}

\correspondingauthor{Marco Gullieuszik}
\email{marco.gullieuszik@inaf.it}

\begin{abstract}
Star-forming, \ha-emitting clumps are found embedded in the gaseous tails of galaxies undergoing intense ram pressure stripping in galaxy clusters,
so-called jellyfish galaxies. These clumps offer a unique opportunity to study star formation under extreme conditions, in the absence of an underlying disk and embedded within the hot intracluster medium. Yet, a comprehensive, high spatial resolution study of these systems is missing.
We obtained UVIS/HST data to observe the first statistical sample of clumps in the tails  and disks of six jellyfish galaxies from the GASP survey;
we used a combination of broad-band (UV to I) filters and a narrow-band \ha filter.
HST observations are needed to study the sizes, stellar masses and ages of the clumps and their clustering hierarchy. These observations will be used to study the clump scaling relations, the universality of the star formation process and verify whether a disk is irrelevant, as hinted by jellyfish galaxy results.
This paper presents the observations, data reduction strategy, and some general results based on the preliminary data analysis. The UVIS high spatial resolution gives an unprecedented sharp view of the complex structure of the inner regions of the galaxies and of the substructures in the galaxy disks. We found clear signatures of stripping in regions very close in projection to the galactic disk.  The star-forming regions in the stripped tails are extremely bright and compact while we did not detect a significant number of star-forming clumps outside those detected by MUSE.
The paper finally presents the development plan for the project.

\end{abstract}

\keywords{
  galaxies: general ---
  galaxies: clusters: general ---
  galaxies: star formation
  galaxies: evolution ---
  ISM: clouds
}

\section{Introduction} \label{sec:intro}

Understanding the physical conditions that lead to the formation of
new stars, and conversely to the halting of the star formation
activity, is central for astrophysics. Galaxy disks are the usual
cradle for star formation (SF); this is a hierarchical process traced by
star-forming regions, dubbed "clumps", 
which are ubiquitous in star-forming
galaxies. High-z galaxies are dominated by bright
clumps, which are larger and more massive than in the local Universe
\citep{elme+2007,fors+2011,guo+2018,zane+2019}, although spatial resolution is clearly an
important factor for the determination of clump properties
as shown by the analysis of lensed high-z galaxies
\citep{cava+2018,mest+2022}. At low-z, HST studies have been fundamental to
obtain a rich panorama of star-forming clumps in galaxy disks from a
number of surveys like e.g. LARS \citep{mess+2019}, LEGUS
\citep{calz+2015}, DYNAMO \citep{fish+2017}.

The SF activity is strongly influenced and can be even
halted by a number of processes, some of which are directly
related to the environment in which the galaxy resides. Ram pressure
stripping \citep[RPS;][]{gunn+1972}, i.e. the removal of interstellar
gas from the disk of star forming galaxies due to the hydrodynamical interaction with the hot intergalactic
medium, is one such process and it is believed to have a strong impact on
galaxy populations in dense environments such as galaxy groups and,
especially, clusters.

Our knowledge of the consequences of RPS on SF activity has
greatly advanced in the last few years.
It has been established that
during the first phase of stripping, SF is enhanced in
galaxy disks, both on kpc-scales and on galaxy-wide scales: most
galaxies undergoing \rr{intense} stripping lie above the star formation rate
(SFR)-stellar mass relation of normal galaxies \citep{vulc+2018, bose+2022}. The
quenching sequence in stripped galaxies has been directly observed,
with outside-in quenching beginning in the external regions of the
disk \citep{owers+2019,gaspIV,bose+2020} leading to \ha\ truncated disks
\citep{koop+2004,frit+2017} and then to post-starburst/post-SF disk
galaxies \citep{gaspXXIV,werl+2022}.

One of the most striking results about the RPS-SF connection is the 
discovery of large numbers of \ha-emitting clumps in the tails of
stripped gas, up to 100-150kpc away from the disk
\citep{foss+2016,merl+2013,cons+2017,gaspXIII}. SF in stripped tails was
seen by some hydrodynamical simulations, though they lack the
capability to predict the characteristics of SF 
complexes \citep{kapf+2009,tonn+2012,roed+2014,lee+2022}.  Observationally, \ha\ clumps,
as well as inter-clump diffuse \ha\ emission, are common in the tails
of so-called ``jellyfish galaxies'', whose long \ha\ tails
are due to intense ram pressure in the central regions of galaxy
clusters where they move at high speed with respect to the intracluster medium (ICM)
\citep{jaff+2018,gaspXXI}.

Recently, the study of jellyfish galaxies has moved from a few
individual cases to statistical samples that have unveiled a number of
results \citep{bose+2018,jach+2019,gaspXIII,pogg+2019b,more+2018,more+2020,more+2020ApJL}: a)
the $\rm H\alpha$ emission of the clumps in the tails is mostly
powered by stars formed in situ,
without an underlying galaxy disk. 
b) the diffuse emission in the tails is due to a combination of
photo-ionization by young stars and heating \rr{(via shock heating or heat conduction)}, the latter most probably
due to mixing of stripped gas with the hot ICM.
c) following the discovery of large amounts of molecular gas in the
stripped tails obtained with single dishes, molecular gas clumps have
now been directly observed with ALMA.
d) the properties of $\sim 500$ H$\alpha$ clumps in jellyfish tails
have been studied at 1 kpc-resolution, finding associated stellar
masses between $10^5$ and $10^8 M_{\odot}$ (similar to high-z
clumps). At these resolutions, the tail clumps seem to follow scaling
relations
($\rm H\alpha$ luminosity vs gas velocity dispersion etc)
similar to disk clumps.

The emerging picture is that gas stripped by ram pressure finds itself
embedded in the hot intracluster plasma but manages to collapse and
form new stars. The light of these young stars is directly observable
in the UV \citep{smit+2010,gaspUVIT,geor+2023}. Depending on the galaxy, stars formed in
the tails represent between a few \% and 20\% of the total ongoing SFR
of the system disk+tail \citep{gaspXIII,werl+2022}.
Overall tail clumps form stars at a higher rate than clumps in the disk with the same stellar mass density;
however, if only the mass formed in the last 100 Myr is considered, the differences are reconciled, suggesting that the local mode of SF is similar in the disks and in the tails on timescales of 100 Myr \citep{gaspXXX}.

One crucial missing piece of the puzzle is a high spatial resolution
study of star-forming clumps in jellyfish galaxy tails and disks. So
far we have no handle on the sizes of a significant number of tail
clumps, and cannot study all the other relevant quantities (stellar
masses and ages, above all) on scales $<$1 kpc.  \cite{cramer2019}
presented a sub-kpc scale study based on HST data of star-forming
clumps in the D100 jellyfish tails; D100 is a low-mass galaxy in the
Coma cluster that has yielded the size and age of three $\rm
H\alpha$-emitting tail clumps and sizes for other 7 candidates.
Another sub-kpc scale study presenting the properties -including the
size- of star-forming clumps in RPS galaxies is the one presented by
\cite{bose+2021}; this study however is focused on IC~3476, a low mass
galaxy in the Virgo cluster that hosts a very low SF activity in the
tail, much lower than typical SF of GASP galaxies \citep{gaspXXI}.

In this paper we present HST observations of 6 spectacular jellyfish galaxies
from the GASP survey \citep{gaspI}.
GASP has obtained MUSE@VLT integral field spectroscopy for 114 galaxies at $0.04 < z < 0.07$,
including 94 ram pressure stripping candidates. Among these, a large number of jellyfish
galaxies -defined as those with the stripped gas tail at least as long as the galaxy stellar disk diameter-
were found. GASP MUSE data are complemented by 
observing campaigns with JVLA, APEX and ALMA, MeerKAT, LOFAR, UVIT,
and Chandra X-ray data to probe all the gas phases and galaxy components.
GASP is providing a substantial contribution
to our understanding of gas stripping processes and RPS in general;
however, GASP results are hampered by the spatial resolution of the
observations that is, in the case of MUSE and ALMA data, $\sim1\arcsec$,
which corresponds to $\sim1$ kpc at the redshift of GASP galaxies. As a
consequence, GASP observations sample
only the largest scales of the star forming structures. The
unique spatial resolution of HST overcomes this limitation and
allows us to gather a fundamental piece of information for
understanding RPS and SF processes in general.
The observations presented in this paper probe
the SF process under uniquely extreme conditions: in the
tails, with no galaxy disk serving as cradle, surrounded by a hostile
hot gaseous environment that in principle could disrupt star-forming
complexes heating them either by mixing, thermal conduction or
shocks. Here, there are self-standing regions that seem to suggest
that the SF process -- once initiated -- cares little about its
underlying and boundary conditions. 
\rr{These regions have however features which are uniquely found in RPS systems. Many star-forming regions in the stripped tails
have elongated or head-tail structures known as fireballs \citep{kenn+2014} with a spatial displacement of the UV and \ha emission that probe SF at different timescales. Moreover young stellar clumps are often organized in long filaments. Both these features are observed in our galaxies and will be further discussed in Sect. \ref{sec:disk} and \ref{sec:uvha}.
}
\rr{Another relevant open question about star-forming regions in the RPS tails is whether, or at what scales, they are gravitationally bound or not; this has direct implications for the fate of these system. \citet{gaspXIII} suggest that the stellar aggregates formed in the tails of stripped gas might contribute to the large population of globular clusters or dwarf galaxies observed in nearby galaxy clusters. \citep{cramer2019} however found that the star-forming clumps in D100 are not gravitationally bound and therefore their stars will be dispersed as a diffuse component of the intracluster light. Addressing this open question and determining
the subsequent fate of the clumps requires high spatial resolution observations to reliably assess the size and mass of the clump; this is one of the key goals of our HST observations that will be addressed as a future development of the project.}

In this paper we present:
the HST observations (Sect \ref{sec:obs});
the data reduction procedure and the evaluation of the data quality (Sect. \ref{sec:dataanalysis});
a discussion of some first qualitative results (Sect. \ref{sec:discussion});
the conclusions and a discussion of the development plan of this project and forthcoming papers (Sect. \ref{sec:sum}).

In this paper we adopt the standard concordance cosmology:
$H_0=70$ $\mathrm{km}\,{{\rm{s}}}^{-1}\,{\mathrm{Mpc}}^{-1}$,
${\rm{\Omega }}_{M}\,=0.3$,
${\rm{\Omega }}_{\rm{\Lambda }}=0.7$.

\section{Observations} \label{sec:obs}
Observations were carried out with the
UVIS \rr{channel of the WFC3} camera onboard the HST between October 2020 and April 2021 (GO 16223; PI Gullieuszik).
All the {\it HST}, data used in this paper can be found in MAST: \dataset[10.17909/tms2-9250]{http://dx.doi.org/10.17909/tms2-9250}.
 
Target galaxies were selected from the sample of GASP ram pressure
stripping galaxies \citep{gaspI} with long \ha\ emitting tails.
The selection was based on the number of \ha\ clumps in the tails
detected with MUSE \citep{gaspXIII}. The main properties of the 6
target galaxies are reported in Tab. \ref{tab:targets}.
The papers published so far using available multi-wavelength observations are listed in
the last column; more data (e.g. with MeerKAT, LOFAR, ATCA, and Chandra) are available and will be published in forthcoming papers.

\begin{deluxetable*}{llcccllc}
\tablecaption{
  Target galaxies name (1),
  host cluster (2),
  equatorial coordinates (3 and 4),
  stellar mass (5), galaxy redshifts (6), and host cluster redshift (7),
  and other available data (8).
  \label{tab:targets}}
\tablehead{
    \colhead{galaxy} &
    \colhead{cluster} &
    \colhead{RA (ICRS)} &
    \colhead{DEC (ICRS)} &
    \colhead{$\log M_\star/M_\sun$} &
    \colhead{$z$}&
    \colhead{$z_{cl}$}&
    \colhead{other data}
    }
\decimalcolnumbers
\startdata
  JO175 & A3716   & 20:51:17.578 & $-$52:49:22.01 & 10.50 & 0.0467 & 0.0456 &\\
  JO201 & A85     & 00:41:30.298 & $-$09:15:45.90 & 10.79 & 0.0446 & 0.0557 &a, b, d, i\\
  JO204 & A957    & 10:13:46.830 & $-$00:54:51.14 & 10.61 & 0.0424 & 0.0450 &a, b, e\\
  JO206 & IIZW108 & 21:13:47.402 & $+$02:28:34.62 & 10.96 & 0.0511 & 0.0489 &a, b, c, f, g\\
  JW39  & A1668   & 13:04:07.729 & $+$19:12:38.36 & 11.23 & 0.0663 & 0.0634 &f\\
  JW100 & A2626   & 23:36:25.045 & $+$21:09:02.88 & 11.47 & 0.0619 & 0.0551 &a, b, e, h\\
\enddata
\tablecomments{
  Values are taken from \cite{vulc+2018} and \cite{gaspXXI}.
  The other available datasets are:
  CO from APEX (a: \citealp{more+2018});
  CO from ALMA (b: \citealp{more+2020ApJL});
  \ion{H}{1} from JVLA (c: \citealp{rama+2019}; d: \citealp{rama+2020}; e: \citealp{deb+2020});
  LOFAR 144 MHz continuum (f: \citealp{igne+2022b});
  magnetic field from JVLA (g: \citealp{mull+2021});
  X-ray from Chandra (h: \citealp{pogg+2019b});
  FUV/NUV from UVIT/ASTROSAT (i: \citealp{gaspUVIT}).
}
\end{deluxetable*}

To probe SF in both the tail and the disk we used the broad band filters F275W and F336W
and the narrow band F680N, which includes the \ha\ spectral line at the redshift of the target galaxy.
To subtract the continuum stellar emission from F680N and to probe the visible band light
we also used the broad-band filters F606W and F814W.
We obtained five HST orbits\footnote{An HST orbit is about 52 minutes long.} for each of the 6 target galaxies
(2 orbits in F275W, 1 orbit each in F336W and F680N, and 0.5 orbit each in F606W and F814W)
for a total of 30 orbits. 
For all observations we used a linear dither pattern to cover the gap
between the 2 UVIS chips, and splitting the exposure time in 4 single
exposures for F275W and 3 exposures for all the other 4 filters.
Details are given in Table \ref{tab:obslog}.

\begin{deluxetable}{lDcccc}
  \tablecaption{
  Description of the observations:
  filter name (1),
  number of orbits (2),
  number of sub-exposures (3)
  total exposure time -- 
  it is slightly different for each galaxy
  and here we report the mean value;
  (5) $5\sigma$ magnitude limit representative for point sources, computed using $5\times 5$ pixels regions;
  (5) $5\sigma$ magnitude limit for $1\arcsec\times \arcsec$ regions. The computation of the last 2 columns is described in Sect \ref{sec:noise}.
  \label{tab:obslog}}
\tablehead{
    \colhead{Filter}&
    \multicolumn2c{Orbits} &
    \colhead{N exp}&
    \colhead{exptime} &
    \colhead{$5\sigma$ mag} &
    \colhead{$5\sigma$ mag}\\
    &&&&sec&point src&$1\arcsec \times 1\arcsec$
}
\decimals
\startdata
F275W & 2   & 4 & 5283  & 26.9 & 25.0\\
F336W & 1   & 3 & 2512  & 26.8 & 24.9\\
F606W & 0.5 & 3 & 1038  & 27.4 & 25.4\\
F814W & 0.5 & 3 & 1058  & 26.5 & 24.5\\
F680N & 1   & 3 & 2507  & 26.1 & 24.1\\
\enddata
\end{deluxetable}

Following the latest STScI’s recommendations (Instrument Science
Report WFC3 2020-08), we used post-flash to mitigate the effects of
UVIS CTE degradation. By using the calculations made with the
Astronomer's Proposal Tool (APT) the post-flash was set to bring the
average background level at 20 $e^-$ per pixel to ensure that CTE
losses remain at a manageable level.

\section{Data analysis}\label{sec:dataanalysis}
\subsection{Data reduction and calibration}\label{sec:datared}
We retrieved from the STScI archive the FLC files with the calibrated images (including the CTE correction)
for each of the sub-exposures.
All files were reduced and calibrated with
the CALWF3 code v3.6.0, including the 2.0 version of the CTE algorithm, that was released in April 2021.

While individual FLC images taken with the same filters
are perfectly astrometrically aligned, in most cases there are small
misalignments between images taken with different filters.  This is most
likely due to the low number of stars bright enough in all filters.
To tweak the astrometric solution first of all we combined all
sub-exposures taken with each filter; we took the F606W as an
astrometric reference and selected, for each of the other 5 combined
images, a number of astrometric reference sources in common with the
F606W one, preferentially non saturated stars, but also compact
extragalactic sources.  The target galaxies have long tails of
gas with star-forming regions; these were not used as astrometric
references as their emission
can be displaced in observations at
different wavelengths (see Sect.  \ref{sec:uvha}).  In all cases we
found a sufficient ($> 10$) number of sources evenly distributed in the field of view.  These were then used
to align all combined images to the F606W one using the \textsc{tweakreg}
task; the astrometric solution was then propagated back to the
FLC files using \textsc{tweakback}\footnote{\textsc{tweakreg}, \textsc{tweakback} and \textsc{AstroDrizzle}, that is used for the following step, are included in the \textsc{DrizzlePac} Python package}.

Cosmic ray removal is a critical task from HST data. In our case it is particularly problematic for the narrow-band F680N frames, as the exposure time is relatively long and we have only 3 frames for each galaxy. Standard procedures could not provide a satisfactory result and we therefore adopted a slightly modified approach.
The cosmic ray maps for the F680N were computed by combining 
all the 9 F606W, F680N, and F814W FLC frames, that were taken during the same visit.

All FLC frames taken with each filter were then stacked together using \textsc{AstroDrizzle}.
Following standard recommendations\footnote{https://www.stsci.edu/scientific-community/software/drizzlepac.html}, we resampled the stacked images with a pixelscale of $0.04\arcsec$, and a \textsc{pixfrac} parameter set to 0.8 for the F275W images (for which we have 4 exposures) and to 1 for all the other filters (for which the exposures are 3).
The output images were visually inspected to look for residual cosmic-rays. The inspection was done by comparing the images of each galaxy in the 5 filters; we flagged as cosmic-rays compact, bright and sharp-shaped emitting regions detected in one filter only.

The final stacked images were then corrected for the Milky Way dust extinction
using \cite{schl+2011} reddening maps and the 
\cite{Cardelli1989} extinction curve.

\begin{figure*}[t]
\centering
 \includegraphics[width=0.9\hsize]{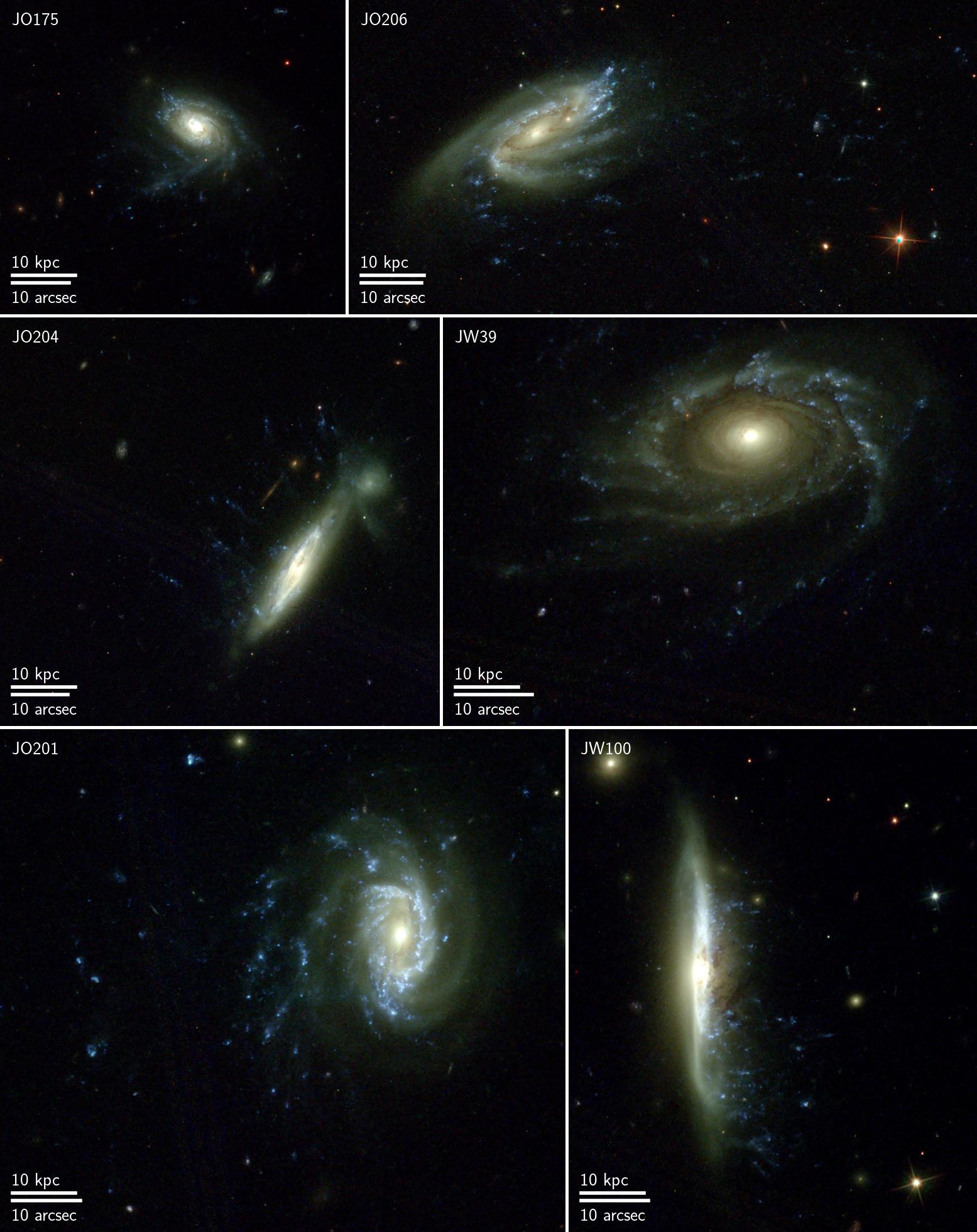}
  \caption{\rr{HST WFC3 images of the six observed galaxies (a description of the data used in each of the  RGB channels is given at the end of Sect. \ref{sec:datared})}.
  The scale in linear (kpc) and angular (arcsec) units are shown on the lower-left corner of each panel.
  All images have been zoomed to have the same scale in linear units (kpc per pixel). The luminosity cuts and stretching function are the same for all galaxies. North is up ad East is left.
  \label{fig:rgb}}
\end{figure*}

The RGB images shown in Fig. \ref{fig:rgb} are obtained using the
F814W and F606W output images for the R and G channels, respectively;
given the large difference in wavelength between the F814W/F606W and the F336W/F275W filters,
for the
B channel we created a pseudo B-band image as:

\begin{equation}
F_\mathrm{B}=
0.25\,(F_\mathrm{F275W}+F_\mathrm{F336W})+
0.5\,F_\mathrm{F606W}
\end{equation}

We then performed a non-local means denoising on the 3 images used in
the 3 channels using the \textsc{scikit-image} python
package\footnote{\url{https://scikit-image.org/}} \citep{scikit}.  All
the images in the 3 channels have been normalized using lower and
upper limits of 0 and $10^{-19}$ \ergscma and an arcsinh stretching
function.
The 6 RGB images were scaled to have the same spatial scale in
kpc/pixel in Fig. \ref{fig:rgb}.

\subsection{Computing the \ha flux}\label{sec:ha}

This section describes the procedure to evaluate the \ha emission flux from the F606W, F680N and F814W observations. First, we assumed that the only line contributing to the F606W and F680N observed fluxes is \ha and that no line
contributes to the F814W flux. Indeed other emission lines (\hb and the [\ion{O}{3}], [\ion{N}{2}] and [\ion{S}{2}] lines being the strongest) are in the spectral range covered by the three filters, but they have a minor contribution in most cases, as we will show in the following.
This assumption can be written as:
\begin{equation}\label{eq:sys_cont}
\begin{split}
F_\mathrm{F606W}=& F_\mathrm{F606W}^\mathrm{cont}+F_{\ha}/ w_\mathrm{F606W} \\ 
F_\mathrm{F680N}=& F_\mathrm{F680N}^\mathrm{cont}+F_{\ha}/ w_\mathrm{F680N} \\ 
F_\mathrm{F814W}=& F_\mathrm{F814W}^\mathrm{cont}\\
\end{split}
\end{equation}
where $F_{\ha}$ is the line flux (in \ergscm) and for each filter $f$:
\begin{itemize}
    \item $F_f$ is the PHOTFLAM calibrated average measured spectral flux density;
    \item $F_f^\mathrm{cont}$ is the average spectral continuum flux density;
    \item $w_f$ is the effective filter width. This was calculated using the {\sc rectwidth} method from the STSDAS package SYNPHOT \citep{synphot}.
\end{itemize}
We also assumed that the spectral continuum flux density is a linear function of the wavelength in the spectral region of F606W, F680N and F814W:

\begin{equation}\label{eq:f680cont}
\begin{split}
F_\mathrm{F680N}^\mathrm{cont}=& a\;F_\mathrm{F606W}^\mathrm{cont}+b\;F_\mathrm{F814W}^\mathrm{cont}\\[.5em]
a= & \frac{\lambda_\mathrm{F814W}-\lambda_\mathrm{F680N}}{\lambda_\mathrm{F814W}-\lambda_\mathrm{F606W}} \\[.5em]
b= & \frac{\lambda_\mathrm{F680N}-\lambda_\mathrm{F606W}}{\lambda_\mathrm{F814W}-\lambda_\mathrm{F606W}}\\[1.5em]
\end{split}
\end{equation}
$\lambda_f$ values have been calculated by creating a flat spectrum in $F_\lambda$ and using the SYNPHOT {\sc effective\_wavelength} method.
The resulting equation is
\begin{equation}\label{eq:fHa}
F_{\ha}=
 411.6 \;F_\mathrm{F680N}
-242.3 \;F_\mathrm{F606W}
-169.4 \;F_\mathrm{F814W}
\end{equation}

As mentioned at the beginning of this paragraph,
the procedure we adopted 
assumes the presence of no other lines except for \ha.
In the following we quantitatively assess
the systematic effects
due to the presence of other emission lines in star forming regions.

\begin{figure}[t]
 \includegraphics[width=\hsize]{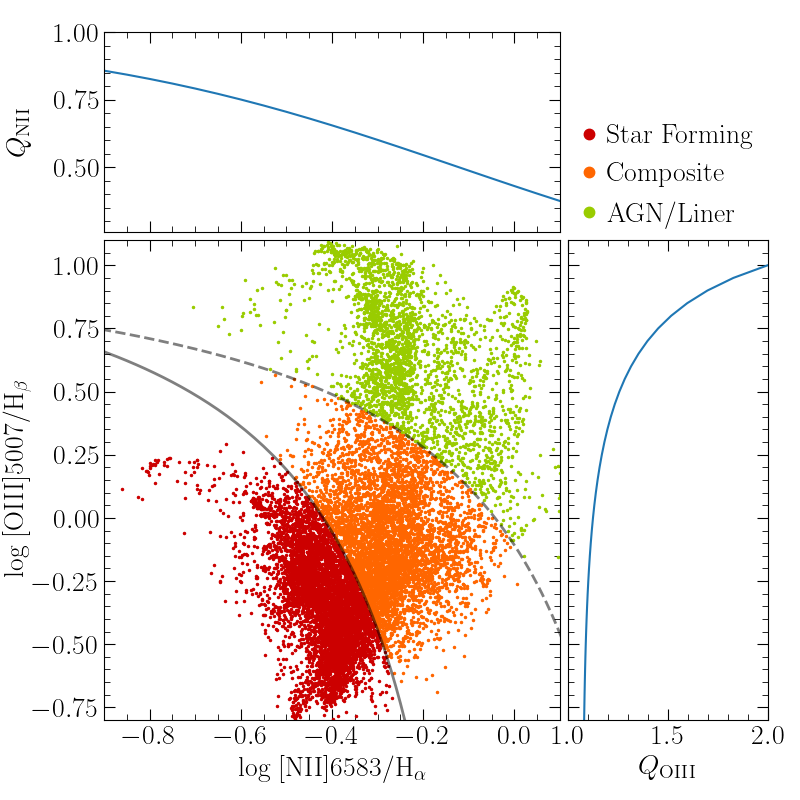}
  \caption{The effect of [\ion{O}{3}] and \hb lines on the \ha flux estimated from our HST data is shown in the right panel; the effect of the two [\ion{N}{2}] on the top one (see text for details).
    The largest panel shows the range of [\ion{O}{3}]/\hb and [\ion{N}2]/\ha line ratios using as a reference the BPT diagram obtained from MUSE observations of JO204 \citep{gaspIV}.
\label{fig:oiii}}
\end{figure}

At the redshift of the target galaxies the [\ion{N}{2}]6548 and [\ion{N}{2}]6583
lines are
included in the spectral range of both the F606W and the F680N
filters.  Since the two lines are very close to \ha, their
contribution is simply summed up to the contribution of \ha.
Therefore Eq. \ref{eq:fHa} actually provides the sum of the flux of \ha and
of the [\ion{N}{2}] lines.
We calculated the ratio $Q_{\mathrm{NII}}$ between the flux of \ha and the total flux of the
three lines for different values of [\ion{N}{2}]6583/\ha 
assuming a fixed line ratio
[\ion{N}{2}]6583/[\ion{N}{2}]6548=3.071 \citep{stor+2000}.
$Q_{\mathrm{NII}}$ is equivalent to the ratio between
the actual value of the \ha flux and the value obtained from
Eq. \ref{eq:fHa}.  Results are shown in Fig.\ref{fig:oiii}. For star
forming regions ($\log $[\ion{N}{2}]6583$/\ha \lesssim -0.3$, red
symbols) $Q_{\mathrm{NII}}$ is always larger than 0.6 which means
that estimating the \ha flux using Eq. \ref{eq:fHa} would overestimate
the true flux by less than a factor $\sim 1.6$.
Even for the regions with the most extreme AGN- or Liner-like line ratios the real \ha flux is at least $\sim$ 40--50\% of the estimated value.

Evaluating the effects of \hb and [\ion{O}{3}] is less straightforward.
At the redshift of the target galaxies these lines are in the F606W band and therefore their emission flux contributes to over-estimate the
F606W stellar continuum and consequently also the computed continuum in F680N (see Eq. \ref{eq:f680cont}); this results in a systematic
underestimation of the \ha\ flux computed from Eq. \ref{eq:fHa}.
To quantify this effect, we used synthetic
spectra with different line ratios to compute the fluxes
in the UVIS photometric band and then we compared the results of
Eq. \ref{eq:fHa} with the value of the input \ha flux. Both input
spectra creation and synthetic photometry computation were carried out
with SYNPHOT.
For the model spectra a continuum described by a linear
function of the wavelength was adopted; we made quantitative tests to verify
that the shape of the continuum has negligible effects on the results.
\hb, \ha, and the [\ion{O}{3}] doublet at 4959 and 5007
\AA\ emission lines were modeled using gaussian
profiles\footnote{The actual shape of the line profile is negligible for this analysis,
as it is based on synthetic photometry on bands much wider than
the line profiles.}.
We adopted a fixed line ratio of 3.013 for [\ion{O}{3}]5007/[\ion{O}{3}]4959 \citep{stor+2000} and 2.86 for \ha/\hb
and created a series of synthetic spectra for different values of the [\ion{O}{3}]5007/\ha ratios. We then used 
SYNPHOT to compute the fluxes in the UVIS photometric bands which were then used to evaluate the $F_{\ha}$ from 
Eq. \ref{eq:fHa}; the ratio between 
the input \ha flux and the resulting value is reported as
$Q_{\mathrm{OIII}}$ in Fig.\ref{fig:oiii}.

As expected, the \ha\ flux is always underestimated ($Q_{\mathrm{OIII}}>1$);
when the [\ion{O}{3}] emission is weak the effect is dominated by the \hb\ emission and it is
$\sim5\%$ which is negligible considering all the sources of uncertainty.
In general, for star forming regions (log [\ion{O}{3}]5007/\hb)$\lesssim 0.25$, see central panel in 
Fig.\ref{fig:oiii})
$Q_{\mathrm{OIII}}$ is smaller than 1.2 which means that \ha\ is never underestimated by more than $\sim15\%$.

\rr{The throughput of the F680N filter is 
essentially constant between 6800 and 7000\,\AA\ with variations of a few percent;
at 7040\,\AA\ its value is decreased by 10\% and at redder wavelengths it drops rapidly.
This might be an issue only for JW39, 
which is the galaxy at the highest redshift (see Tab. \ref{tab:targets});
using results from GASP observations with MUSE we can safely assume that there should not be any
\ha or [\ion{N}{2}] emission at wavelength longer than 7040\,\AA\,. We therefore conclude that the dependency of the filter throughput on wavelength has no significant effects on the \ha flux estimate.}

We finally note that our considerations do not take into account dust effects;
as dust attenuates \hb\ and [\ion{O}{3}] more than \ha\ emission it would therefore decrease the $Q_{\mathrm{OIII}}$ value;
as a consequence, the values discussed above are upper limits as any SF regions would have a non-negligible dust extinction.

We can therefore conclude that the main source of uncertainty in our method to derive the \ha\ flux is due to
the contribution of the [\ion{N}{2}] lines. All other emission lines play a second order role that would in any case
act in opposite direction to the one of [\ion{N}{2}]; their contribution
would therefore reduce the systematic effect due to the [\ion{N}{2}] lines.

\subsection{Background variation, noise and detection limit}\label{sec:noise}

\begin{figure*}[t]
 \includegraphics[width=\hsize]{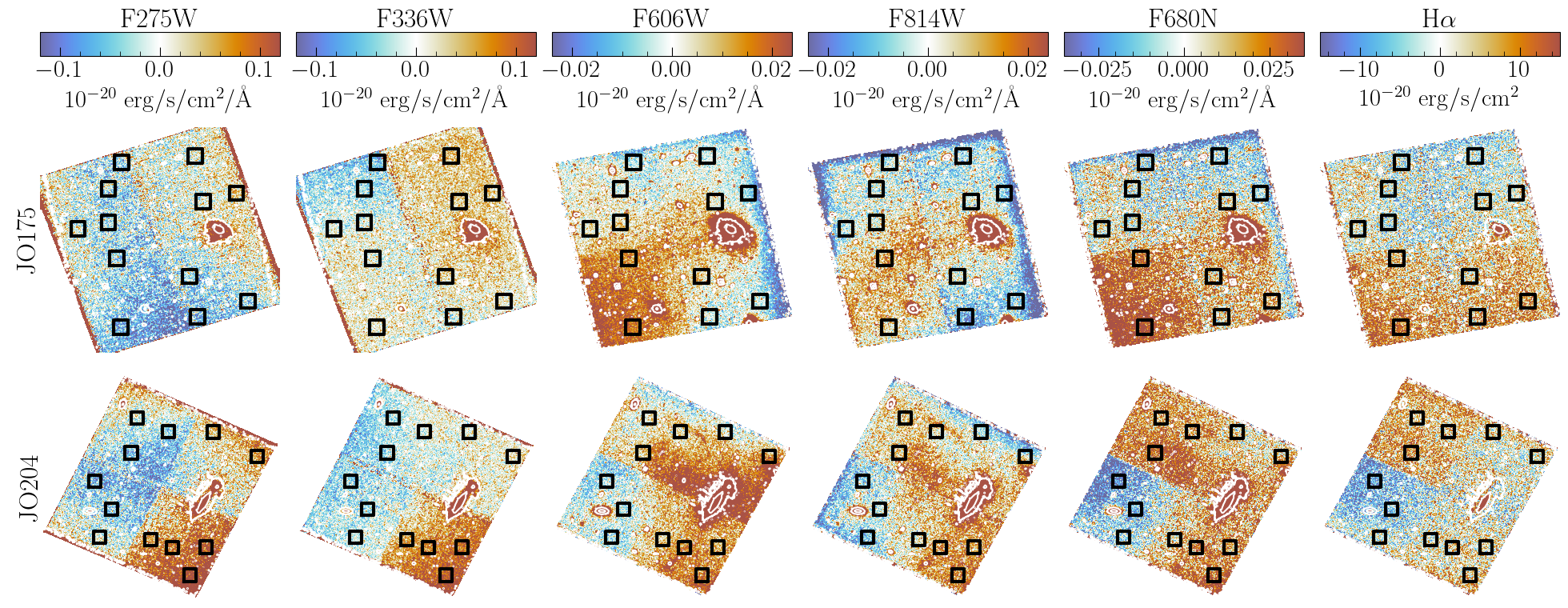}
 \caption{Images in the 5 UVIS filtes and the \ha\ emission map for two target galaxies.
   Images were smoothed and convoluted with a gaussian kernel to reduce the noise on small spatial scales
   and to highlight large scale background variations.
   The white contours are isophotes from the F606W image shown to visualize the stellar sources as a reference.
   The black squares show the 12 regions that were used to evaluate
   the statistical properties of the images.
   \label{fig:bkg}}
\end{figure*}

\begin{figure}[t]
 \includegraphics[width=\hsize]{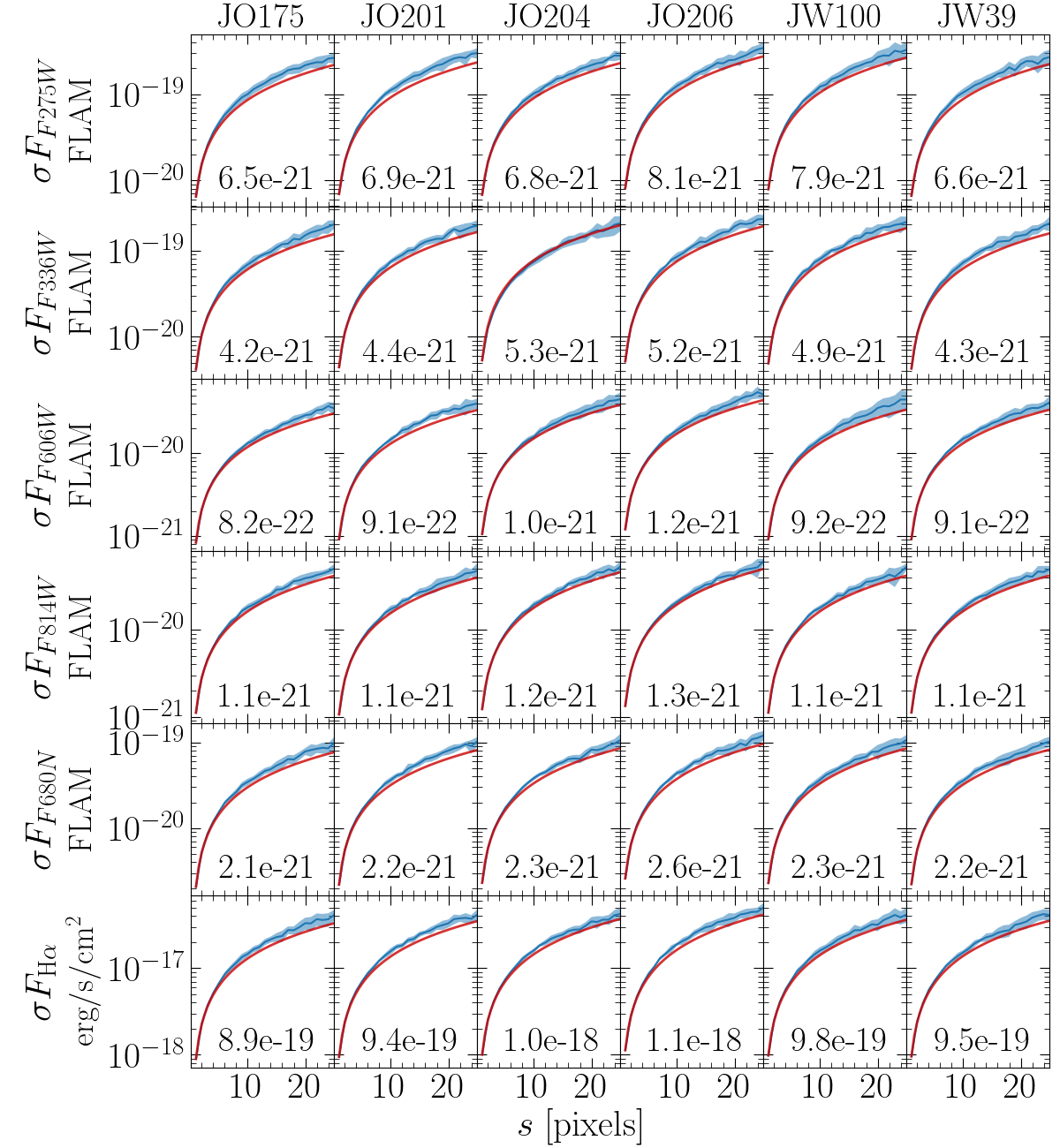}
 \caption{Mean value and dispersion (blue area) of the noise level at different scales
   computed in 12 different regions on each image.
   The red lines are predictions using the formula in \cite{case+2000}.
   The values of the noise on 1 pix scales $\sigma_1$ are reported in each panel.
   Noise values are in \ergscma for the images in the 5 filters (panels in the top 5 rows) and in \ergscm for the \ha emission map (bottom row).
\label{fig:noise}}
\end{figure}

This section presents a statistical analysis of our images and 
assessment of the noise level and the magnitude limit of our observations.

By construction, the average background in the final images is zero, but small-amplitude
residuals are present on large spatial scales.
These
are shown in Fig. \ref{fig:bkg} for two galaxies as an example.
With the aim of
enhancing large-scale structures and minimizing the local noise,
in Fig. \ref{fig:bkg}
we binned $10\times10$ the DRC images and the results were divided by 100;
the values in the resulting image can be therefore interpreted as 
\rr{spatially-averaged image with the same intensity scale as the original image}.
\rr{We then smoothed the image by convolving it with a 2D Gaussian kernel with a standard deviation of 3 pixels}. 
The rightmost panels in Fig. \ref{fig:bkg} show also results for the \ha\ emission maps obtained as described in Sect. \ref{sec:ha}.

First of all we note that in some of the images there are discontinuities corresponding to the
edges of the 2 UVIS chips and of the two readout amplifiers in each of
them.
Besides these, all other large scale patterns are different from one image to the other;
they might be due to a combination of uncertainties in the image
reduction and calibration and/or to stray light components.  We note
that background variations are in all cases very small, of the order
of $10^{-21}$ and $10^{-22}$ \ergscma\ for the images in the two UV
and the three visible filters, respectively. As we will show in the
following, this is smaller than the
average value of the local $1\sigma$ noise in the images. A further
characterization of the background variations and investigation of its
origin are therefore beyond the scope of this paper as they do not significantly affect any of the conclusions of this work nor of the photometric measurements that will be used in follow-up analyses.

The drizzling procedure induces correlated noise \citep{drizzlepac}
which was evaluated by carrying out a statistical analysis on regions
of different size, from 1 (40 mas) to 25 pixels ($1\arcsec$).
First of all, we selected for each galaxy 12 regions of $300\times300$ pixels with
no bright sources (see Fig. \ref{fig:bkg}).
Since some faint --not detected-- sources might be
present, for each region, the background mean brightness value $m$ and the rms noise
$\sigma$ were estimated with an iterative procedure.
First guess values
were estimated as the mean and the standard deviation of the
counts in the region. We then fitted a Gaussian function to the values lower than
$m+\sigma$. The $m$ and $\sigma$ values derived by the fitting
procedure were then used to update the selection procedure and the
fitting procedure was repeated to obtain a final value for the mean
and rms noise of the counts.  We then re-binned each of the 12 cutout
images \rr{(one for each of the 12 regions)} using binning factors from 2 to 25,
to estimate the statistical properties of the images on scales up to 1\arcsec (25 pixels).

The standard deviation of the count rate in each of the 12 regions for
all re-binning factors gives the noise on the corresponding spatial
scale.  The 12 noise values are consistent with each other indicating,
as expected, that the noise level of the background is substantially constant
across the images.  For each spatial scale we calculated
the average and the standard deviation of the noise values obtained for each
of the 12 regions; these are shown in Fig. \ref{fig:noise} as blue
shaded area. The figure shows also
predictions for the correlated noise using the formula from \cite{case+2000}; for images
drizzled with a {\sc PIXFRAC} parameter equals to $p$, the noise on
scales corresponding to $N$ pixels is:
\begin{equation}
\begin{split}
\bar{\sigma}&=\sigma(1)/(1-p/3) \\
\sigma(N)&=\bar{\sigma}\,N\,(1-p\,N/3)
\end{split}
\end{equation}
for our images $p=0.8$ for F275W observations, 1.0 for all other filters (Sect. \ref{sec:datared}).
These relations are shown as a red line in Fig. \ref{fig:noise} and show a good agreement with our data;
the noise model from \cite{case+2000} underestimates the observed values in particular at large spatial scales.
This is a well known fact that is commonly associated
to a combination of the presence of very
faint sources and small irregularities in the background, as was already found by \cite{case+2000}.

These results were used to compute the magnitude limit of our
observations. For each filter, we computed the mean values of the
noise levels in Fig.\ref{fig:noise} and we converted them into AB
magnitudes using the UVIS zero-points;
to estimate the detection magnitude limit for point sources
we used the noise computed on $5\times5$ pixels regions, corresponding to the values generally used on exposure time calculator;
a magnitude limit more representative for extended sources was computed from the values obtained for $1\arcsec \times 1\arcsec$.
The results are shown in Table \ref{tab:obslog}; they are in good agreement with the values obtained with the UVIS exposure time calculator.
For the \ha flux we obtained a $1\sigma$ detection limit of 
$7\times10^{-18}$ \ergscm for point sources and 
$4\times10^{-17}$
\ergscm\,arcsec$^{-2}$ for diffuse emission.

At the redshift of the target galaxies  ($z\sim0.05$) the \ha point source detection limit corresponds to 
a luminosity $L_{\ha}=4\times10^{37}$  \ergs. This value is very similar to the luminosity of the faintest clumps detected for GASP galaxies with MUSE \citep{gaspXIII}. We note however that for clumps larger than the UVIS resolution ($\sim 70$pc FWHM) the star forming clumps can not be approximated by point-like sources; in this case their flux would be spread on a larger area and consequently the detection limit would be brighter than the one estimated for point-like sources.

\section{Discussion}\label{sec:discussion}

The spatial resolution of HST observations gives an unprecedented
detailed view of the target galaxies; previous observations of these
galaxies at visible wavelengths with VLT/MUSE \citep{gaspI} and in the UV
with ASTROSAT/UVIT \citep{gaspUVIT} have spatial resolution of the order of 1\arcsec; HST UVIS data are at least 10 times better than this, opening
a completely new window to probe sub-kpc scale structures on GASP
galaxies.

\subsection{Central regions of the galaxies}

This section focus on the central region of the 6 target galaxies; the UVIS data allow us to study with unprecedented
high resolution the source of the
bright central emission and to complement the available MUSE spectroscopy to better constrain its nature.
Results from the GASP survey showed that
JO201, JO204, JO206 and JW100 galaxies host an AGN \citep{rado+2019};
JW39 central region has a LINER-like spectrum \citep{pelu+2022}; finally ionization in the central region of
JO175 is dominated by SF activity
\citep{gaspXIII,rado+2019}.

\begin{figure*}[!t]
\centering
\includegraphics[width=\hsize]{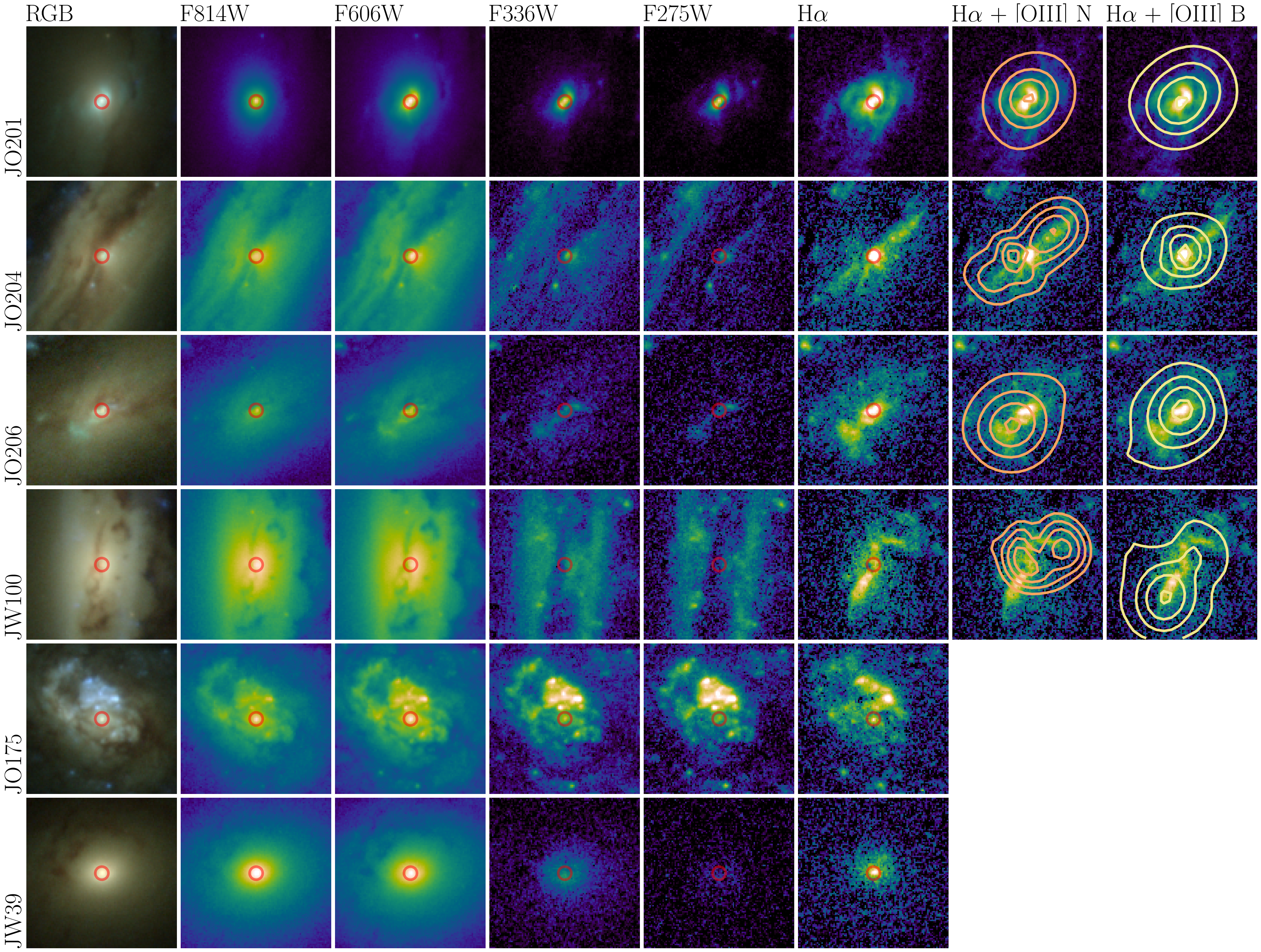}
\caption{RGB image, \ha\ emission map, and broad-band images of the central $4\arcsec\times4\arcsec$ of the 6 target galaxies. The red circle marks the center of the galaxies. 
The RGB images in the leftmost column are obtained as the other ones presented in this paper but using different luminosity cuts to better visualize the bright central regions.
The background image in the two rightmost panels is the \ha\ emission map; the contours show the flux of the broad (light yellow) and narrow (light orange) components of the [\ion{O}{3}]5007 detected with MUSE \citep{rado+2019}.
An interactive visualization of the data presnted here is available at \url{https://web.oapd.inaf.it/gullieuszik/hst_gasp_centers} and in the online journal.
\label{fig:center}}
\end{figure*}

The central regions of the galaxies are shown in Fig. \ref{fig:center}.
A large number of very compact sources are detected in the central region of JO175;
they are bright both in the UV and in the \ha\ maps and faint in F814W; we can therefore conclude they are
young star forming regions, confirming the results obtained from MUSE \citep{gaspXIII,rado+2019}.
JW39 emission is dominated by a source with a regular elliptical morphology at the center of the galaxy; it is clearly detected in \ha\ and in F336W 
and extremely faint in the F275W; indeed among the 6 central sources it is the one with the reddest F275W-F336W color;
the most plausible scenario is that the central region is obscured by dust and older than
the central regions of the other galaxies.

All the other 4 galaxies host an AGN; 
the [\ion{O}{3}]5007 line is therefore expected to be extremely strong in the central regions
and therefore the \ha flux we computed could be significantly under-estimated
as discussed in Sect. \ref{sec:ha}; we note that
none of the conclusions of this paper, nor any planned
analysis will be based on the \ha flux of the regions dominated by AGNs.

The morphology of the central regions for these galaxies varies significantly at the different wavelengths (Fig. \ref{fig:center}).
Multiple dust lanes are present in the nuclear regions, in particular in JO204, JO206 and JW100: this is often observed in HST images of type-2 AGN \cite[see e.g.][]{keel+2015,ma+2021}. In particular, \citet{keel+2015} attributed the presence of irregularly distributed dust lanes, similar to  those observed here, to ongoing or past interaction processes.

HST observations also reveal the detailed morphology of the bright \ha\ emission already observed with MUSE, 
that was proved to be \rr{predominantly} ionized by the AGN \rr{based on the analysis of the emission line ratios in MUSE \citep{rado+2019}}.
MUSE data  also showed that emission lines in JO201, JO204 and JW100 clearly present at least two components in the line profiles,  one {\em broader} ($\sigma_v \sim 200-500$ km s$^{-1}$)  and the other {\em  narrower} ($\sigma_v < 200$ km s$^{-1}$).  \rr{Both components can be related to outflows, which in  \citet{rado+2019} were  defined  as those cases where the line velocities  deviate significantly from the rotational field traced by stellar velocities measured at the same positions \citep[see also e.g.][]{2020MNRAS.498.4150D}}.
In Fig. \ref{fig:center} we overlay on the HST \ha\ images the contour maps of the  [\ion{O}{3}]5007 flux in these two components.
This allows us to compare the HST \ha\ properties with the analysis of outflow properties done in \cite{rado+2019}, to which we refer for more details. 
In JO201  both  [\ion{O}{3}] components overlap well with the \ha\ central, more compact emission; this agrees with the interpretation that in JO201 the outflow orientation is close to the line of sight.
In JO204 the \ha emission shows a more  extended structure:  the broader  [\ion{O}{3}] component overlaps with the central emission,  the narrower one is peaked at two opposite positions along  the \ha\ emission; 
both components were associated to the outflow.
Similarly, in JW100 there are two [\ion{O}{3}]  components of similar width \rr{($\sigma_v < 200$ km s$^{-1}$) emitted by distinct regions along the \ha emission and merging into a double peaked profile in the inner regions}, which were also associated to an outflow.
Finally, [\ion{O}{3}] shows a very faint broader component in the central MUSE spaxels of JO206, and \cite{rado+2019} concluded that it was not possible to detect a meaningful outflow component: we notice however that this component overlaps well with the \ha\ peak.
The dominant, narrower component still follows the \ha\ emission, but there is an offset  between the peaks of the two maps. 
\rr{To summarize,  though both the broader and narrower components may be related to outflows and the \ha\ maps include the contribution from both,  it can be seen that the emission from the broader component is peaked on the central \ha\ emission, thus confirming that it is mostly emitted by the inner nuclear regions. The narrower components are instead  coincident with a more extended \ha\ emission.}

\subsection{Galaxy disk and stripped gas in the inner regions}\label{sec:disk}
\begin{figure*}[!t]
\centering
\includegraphics[width=\hsize]{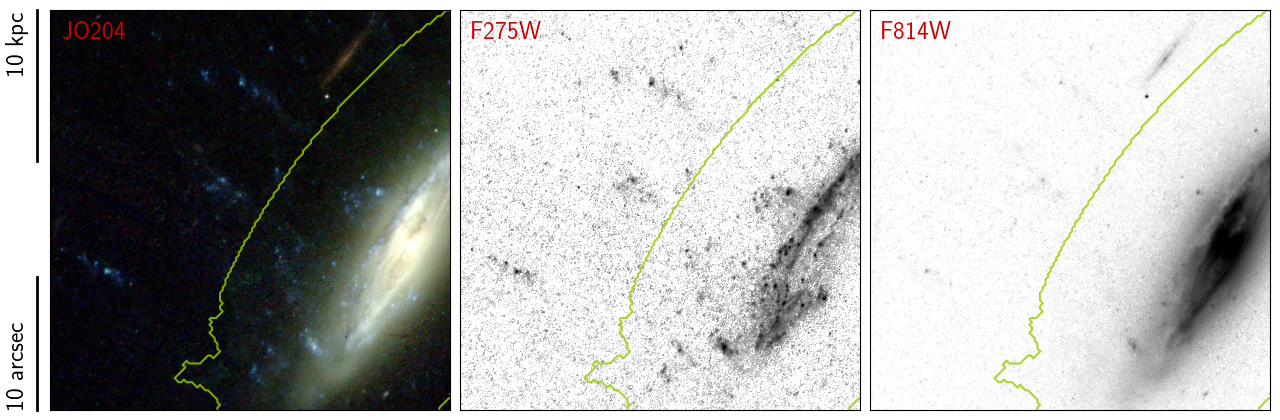}\\
\includegraphics[width=\hsize]{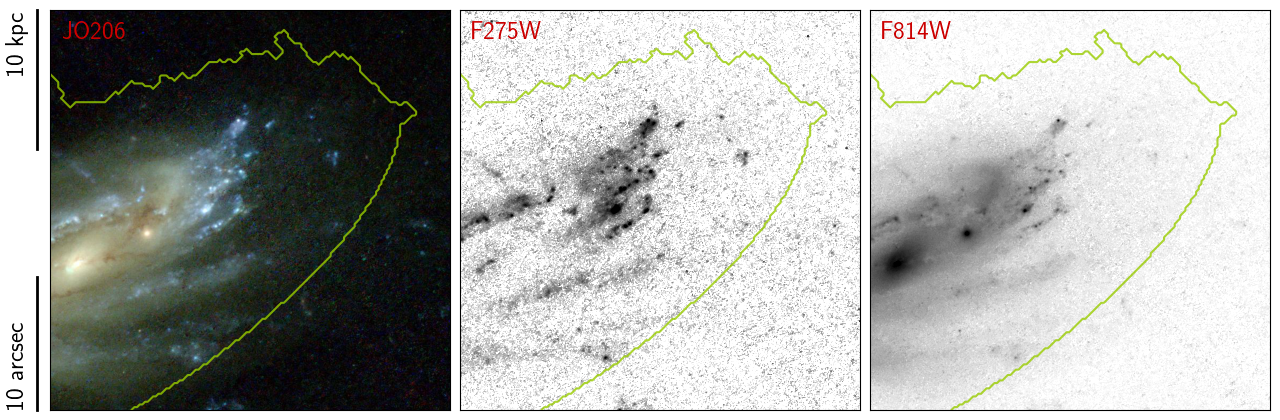}\\
\includegraphics[width=\hsize]{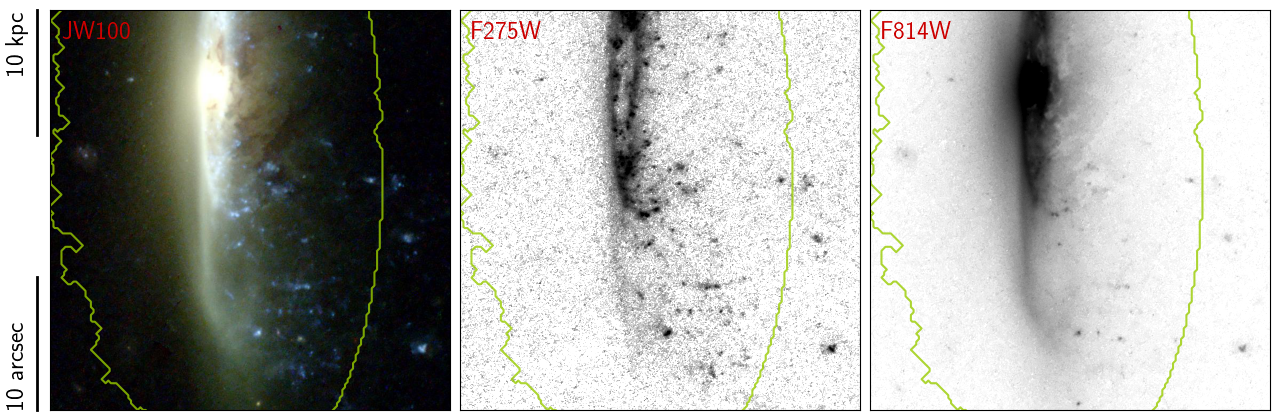}
\caption{RGB, F275W, and F814W images of JO204, JO206, and JW100.  The
  green line is the galaxy disk contour derived from MUSE observations
  \citep{gaspXXI}.
  \label{fig:rgb_uv_814}}
\end{figure*}

The high potential of the diagnostic power of UVIS
observations can be appreciated in Fig. \ref{fig:rgb_uv_814} in
which we show a zoom on the galaxy disk of JO204, JO206, and JW100 of
the RGB images from Fig. \ref{fig:rgb}. The same figure also shows the
F275W and F814W images, to probe star forming and intermediate-old
stellar populations.  In each panel we also show, as a reference, the
line derived from GASP data to define the stellar disk and the
stripped gas tail \citep{gaspXXI}.

HST observations provide a much detailed view of the complex structure
of the galaxy disk and of the inner galactic regions; the 1 kpc
spatial resolution of MUSE and UVIT observations does not allow to
clearly resolve and characterize substructures in particular for
galaxies with a substantial inclination in the plane of the sky. As a
consequence, it is also extremely difficult to clearly
disentangle structures belonging to the disk and to the stripped tail;
previous GASP works therefore adopted a conservative approach to define SF regions
that can be safely classified as being formed in the stripped gas
tail.  This was done by considering the stellar continuum emission and
using the isophote corresponding to a surface brightness $1\sigma$
above the average background; the isophote is not symmetric due to RPS
and hence an ellipse was fitted to the undisturbed side of the
isophote; this ellipse was used to replace the isophote on the
disturbed side \citep[for details see][]{gaspXXI}.  The resulting
contour line is shown in green in Fig. \ref{fig:rgb_uv_814}.
HST observations allow us to go beyond this simple approach and they provide
robust clues on how RPS affects the inner region of galaxies.  Figure
\ref{fig:rgb_uv_814} shows bright regions in the disk that are
particularly bright in the UV and barely visible -or not visible at
all- in the F814W image; they hence stands out as bright and blue sources in
the RGB images; most of them have also \ha\ emission associated with
the UV bright emission.  We also note that they are organized in filamentary
structures aligned in the direction of the tail. For these
reasons we can safely
conclude that the observed bright regions are young stellar clumps that are formed in gas
stripped from the stellar disk by ram pressure. Being so close
to the galactic disk, we can not say whether they are still
gravitationally bound to the galaxy --and will eventually fall back in
the galaxy disk-- or not and hence they will be completely stripped
and lost in the ICM.
A further investigation and characterization of these regions in the disk that
are affected by gas stripping is presented in \cite{paper2}.

Many SF regions are organized in filamentary structures, in
agreement with cloud-crushing simulations of cold clouds in a hot wind; these have been able to produce long tails of cold, dense gas that are about the cloud width and extend for tens of cloud radii when the radiative cooling time is shorter than the cloud destruction time \cite[e.g.][]{gronk+2018,abru+2022,tan+2022}.
Recent simulations have found star formation within these streams from individual dense $\sim$ 100 pc clouds (Tonnesen \& Smith, in prep).

\subsection{Star-forming clumps and diffuse emission}
A major point that previous GASP observations could not address directly
concerns the nature of the diffuse \ha emission observed in GASP galaxies \citep[][and refs therein]{tomi+2021GASP32,tomi+2021GASP35}; in the tail, the diffuse emission (defined as the \ha component outside compact clumps) is found to be on average 50\% and up to 80\% in some galaxies \citep{gaspXIII}.
Diagnostic BPT diagrams based on [\ion{N}{2}]/\ha line ratio indicate that the dominant ionization source of the diffuse emission is SF; however 
MUSE data shows that other mechanisms are at play, like mixing, shocks, and accretion of inter-cluster and interstellar medium gas \citep{tomi+2021GASP32,tomi+2021GASP35}.
MUSE observations could not firmly establish whether the diffuse emission powered by SF was due to ionizing radiation escaped from the star forming clumps detected by MUSE or from a population of smaller and undetected star forming clumps that are hiding within the diffuse \ha emission in the tails.

In \citet{paper2} we present a thorough characterization of the star forming clumps detected from the UVIS observations presented in this paper. These are identified either in UV or \ha
with luminosity down to $\sim10^{36}$ \ergsa in F275W and $\sim10^{38} $\ergs in \ha; these values are very close to the detection limit computed in Sect. \ref{sec:noise}. 
We did not detect any significant population of compact sources in UV nor in \ha in the tails outside of the star forming clumps detected with MUSE \citep[see][]{gaspXIII}. Hence we can safely conclude that the ionizing source of what was defined as diffuse ionized emission from MUSE data is not in-situ SF in clumps brighter than the detection limit of our HST observations.

Another interesting result from \cite{paper2}
is that the sizes of the clumps in the tails measured from the HST observations \rr{are} generally smaller than what was estimated in \citet{gaspXIII} from MUSE observations using the luminosity-size relation for \ion{H}{2} regions from \cite{wisi+2012}. However, we found that at a given size, tail clumps are $\sim 10$ times brighter than the \ion{H}{2} regions observed by \cite{wisi+2012}.
The paucity of very large star-forming clumps (larger than a few 100 pc) and/or the compactness of the star forming clumps in the tails might be connected with the peculiar physical condition in the ram pressure stripped gas which might affect the SF process;
in principle, the collapse of molecular clouds and the SF processes
could be influenced by thermal conduction from the ICM; however this effect could be mitigated or even prevented by magnetic fields \citep{mull+2021, igne+2022}.
The effect of the complex interplay between the stripped gas and the ICM would also affect
the turbulence of the gas in
the tails and hence the properties of the clustering hierarchy which should be hence linked to
the environment and its pressure. Future work based on the observations presented here and on the properties of the SF regions detected in \cite{paper2} will study the dependency of SF clustering on local
environment by comparing the size distributions and fractal properties of the SF regions in
the tails and in the disks with those of undisturbed galaxies.

\subsection{UV and \ha} \label{sec:uvha}

\begin{figure}[!t]
\centering
\includegraphics[width=.49\hsize]{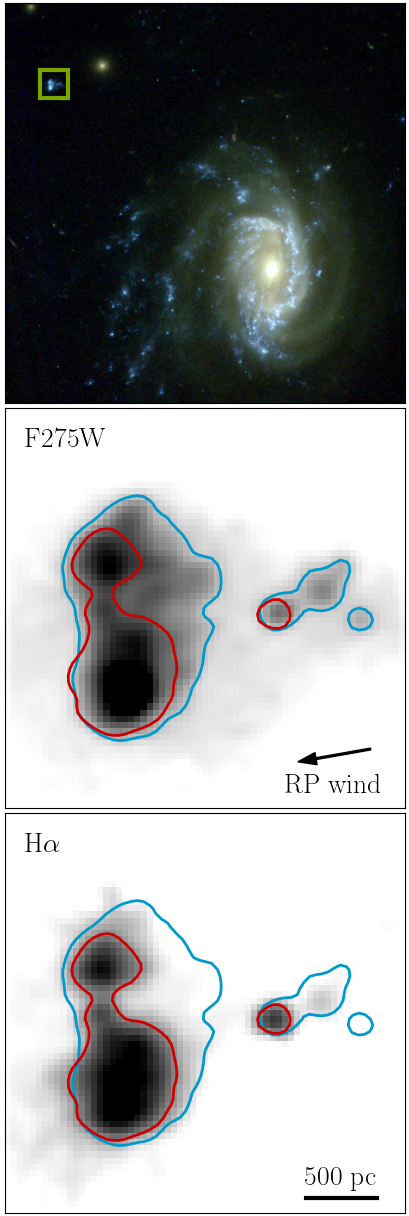}
\includegraphics[width=.49\hsize]{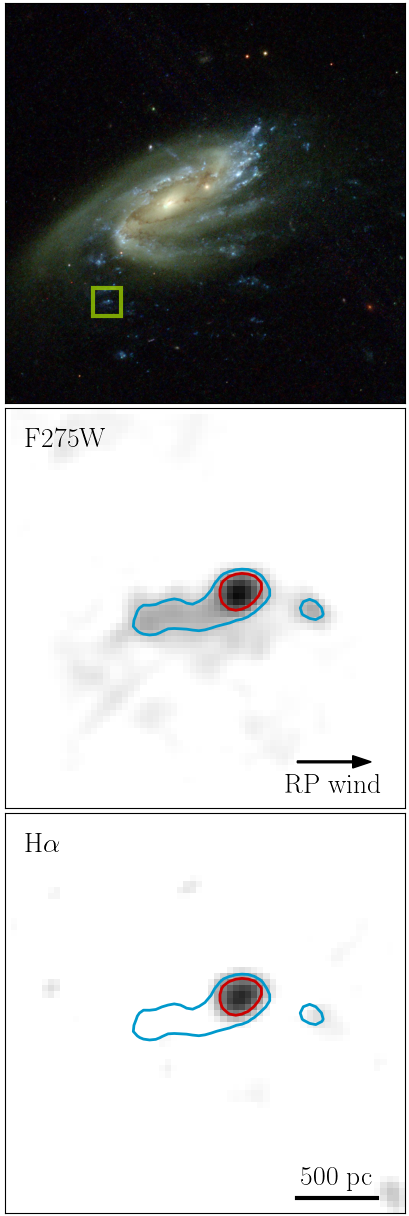}
\caption{Zoom of the F275W and \ha\ images of JO201 (left panels) and
  JO206 (right panels) on two SF regions; their position is marked by
  the green squares in the top panels.  The blue and red lines are
  isophotes of the F275W and \ha\ emission, respectively, \rr{which are over-plotted to highlight the differences in the spatial distribution of \ha and UV emission. The arrows in the central panels suggest the approximate direction of the ram-pressure wind.} 
  \label{fig:uvha}}
\end{figure}

Since the gas in the tail is constantly accelerated by ram pressure it is therefore expected that stellar populations of different ages formed by the same parent gas cloud might be found at slightly different spatial locations;
RPS tails show fireball structures \citep[see eg][]{kenn+2014} with elongated UV emission (tracing SF on timescales $\sim10^8\,$yr) that in some case have head-tail structure
and compact \ha emission (tracing SF on timescales $\sim10^7\,$yr) located on the head -in the direction of the ram pressure wind.
\rr{Numerical simulations predict general alignment of \ha and UV emission, with UV emission extending somewhat closer to the disk and \ha emission extending slightly further from the disk. This partially differs from the $\sim100$ pc displacement between UV and \ha emission that was found, for example, in the tail of IC3418 by \citet{kenn+2014}.
In our data there are some notable examples of this effect, and two of them are shown in Fig. \ref{fig:uvha}. The head-tail structures detected in UV have sizes of a few 100 pc and  the \ha emission is concentrated in compact regions on one side of the extended UV emission, in the downstream direction of the ram pressure wind; the peak of \ha and UV emission are however coincident. This tends to be in better agreement with what is seen in simulations.}

Galaxy-scale simulations generally do not reproduce clear stellar age gradients in the tail, although some of this can be blamed on their inability to resolve the displacements observed in our star-forming clumps \citep[e.g.][]{kapf+2009,tonn+2012,roed+2014}.
In the cloud-scale simulations of Tonnesen \& Smith (in prep), age gradients are not universally seen in the stars formed in stripped clouds.  They find that the faster and denser a wind is, the more likely an age gradient is produced that would be reflected in an observed displacement of UV and \ha.  In addition, more diffuse clouds that are still able to collapse are more likely to show age gradients.  

  A systematic quantitative analysis of the different morphology of UV and \ha emission is beyond the scope of this paper. Ongoing and planned work on these observations include a systematic search and characterization of SF regions and
  a detailed analysis of the stellar populations using SED fitting. Forthcoming papers based on the observations presented in this paper will therefore thoroughly examine the points briefly discussed in this section.

\section{Summary and project development plan}\label{sec:sum}
This paper presents UVIS HST observations of 6 RPS galaxies at redshift $z\sim0.05$ from the GASP survey \cite{gaspI}; they were selected for hosting a large number of star forming clumps in the tails of stripped gas \cite[see][]{gaspXIII}.
Observations were carried out in four broad bands covering a spectral range UV to the I-band (F275W, F336W, F606W and F814W) and a narrow band one (F680N) covering \ha emission at the redshift of the target galaxies.

The main goal of this observing programme is to complement the large dataset collected within the GASP project, which is based on MUSE observations and that was then followed-up with multiwavelength observations with JVLA, APEX and ALMA, MeerKAT, LOFAR, UVIT, and archival Chandra X-ray data to probe molecular and neutral gas, as well as UV and X-ray emission. The main limitation of these data is the spatial resolution, which is of the order of $1\arcsec$ for all observations, corresponding to $\sim 1$\,kpc. The HST observations presented here allow studies of GASP galaxies with unprecedented spatial resolution; this is a critical aspect, in particular to characterize SF regions which are in general small sources with sub-kpc scales.
Moreover, these observations provide deep UV data that would strongly constrain the
properties of young stellar populations, as well as \ha data probing the ionized gas.

This paper presents a general description of the observations and the data reduction process; it also presents some general discussion and results that have been drawn from a preliminary analysis of the data. These can be shortly summarized as follow:
\begin{itemize}
    \item
    We do not detect a significant number of compact \ha or UV sources in \ha emitting regions outside the star forming clumps detected with MUSE.
    This shows that the ionizing source of this diffuse ionized gas component 
    is not in-situ SF in clumps above the UVIS detection limit ($L_{\ha}=10^{38}$\,\ergs at 2$\sigma$).
    \item
    The vast majority of the clumps detected in the stripped gas are not complex structures;
    HST images reveal that nearly all clumps detected with MUSE are single compact and bright sources.
    \item
    Thanks to the extraordinary spatial resolution of HST we found 
    clear signatures of stripping also in regions that are very close in projection to the galaxy disks.
    \item
    There are some examples of tail clumps that show a clear difference in the UV and \ha emission. UV emission is elongated in the direction of the speed of the galaxy in the ICM (and hence of the RPS) while \ha is concentrated on the side of the UV emission opposite to the position of the galaxy. 
    This fireball structure \citep[see][]{kenn+2014} is the effect of 
    the different ages probed by UV and \ha for a star-forming gas cloud that is accelerated by ram pressure.
\end{itemize}

The project based on the HST observations presented here will be developed in a series of research programmes. As already mentioned, a thorough
analysis of the star forming clumps is presented in \cite{paper2}.
With ongoing SED analysis of the star-forming clumps we are analyzing the properties of the stellar populations (ages, star-formation histories and stellar masses) of the star forming regions in the stripped tails and in the galactic;
the sizes and masses of individual clumps are crucial to establish the current nature of the clumps and to investigate whether they  resemble the properties of e.g. globular clusters, UltraCompact Dwarfs or Dwarf Spheroidals and to understand what could be their future evolution. To date, the limited spatial resolution of the available data could provide
only upper limits on individual masses and most probably largely inflated sizes; these can not provide sufficient constraints and are compatible with all the above mentioned hypothesis \citep{gaspXIII}. HST data provide for the first time the accuracy required to reliably assess the nature and fate of the observed clumps.

\acknowledgments We would like to sincerely thank the referee for their constructive comments that helped us improving the quality of our manuscript. 
We warmly thank Jay Anderson, Crystal Mannfolk, and the HST Help Desk staff for the valuable support.
This research is based on observations made with the
NASA/ESA Hubble Space Telescope obtained from the Space Telescope
Science Institute, which is operated by the Association of
Universities for Research in Astronomy, Inc., under NASA contract NAS
5–26555. These observations are associated with program GO-16223.
This research made use of Astropy, a community-developed core Python package for Astronomy (Astropy Collaboration, \citeyear{astropy}).
This project has received funding from the European Research Council (ERC) under the European Union's Horizon 2020 research and innovation programme (grant agreement No. 833824).
and "INAF main-streams" funding programme (PI B. Vulcani).
YJ acknowledges financial support from ANID BASAL Project No. FB210003.

\bibliographystyle{aasjournal}

\software{Python, 
astropy \citep{astropy},
synphot \citep{synphot}}

\end{document}